\documentclass[aps,prd,a4paper,onecolumn,amsmath,showpacs,superscriptaddress,nofootinbib,preprintnumbers]{revtex4-1}

\usepackage{verbatim}
\usepackage[T1]{fontenc}
\usepackage[utf8]{inputenc}
\usepackage[american]{babel}
\usepackage{epsfig}
\usepackage{graphicx,subcaption,caption}
\usepackage{booktabs}
\usepackage{multirow}
\usepackage{dcolumn}
\usepackage{amsmath}
\usepackage{mathtools}
\usepackage{amsfonts}
\usepackage{amssymb}
\usepackage{epstopdf}
\usepackage{bm}
\usepackage{siunitx}
\usepackage{braket}
\usepackage{enumitem}
\usepackage{soul}
\usepackage[table]{xcolor}
\usepackage{color}
\usepackage{transparent}
\usepackage{pifont}

\definecolor{navyblue}{rgb}{0.0, 0.0, 0.5}
\definecolor{royalblue}{rgb}{0.25, 0.41, 0.88}
\definecolor{cadmiumgreen}{rgb}{0.0, 0.42, 0.24}
\definecolor{blue-violet}{rgb}{0.54, 0.17, 0.89}
\definecolor{darkviolet}{rgb}{0.58, 0.0, 0.83}
\definecolor{orange(colorwheel)}{rgb}{1.0, 0.5, 0.0}

\usepackage{hyperref}
\hypersetup{
    colorlinks=true, 
    linkcolor=royalblue, 
    citecolor=magenta}

\newcommand\ee{\end{equation}}
\newcommand\be{\begin{equation}}
\newcommand\eea{\end{eqnarray}}
\newcommand\bea{\begin{eqnarray}}

\renewcommand\({\left(}
\renewcommand\){\right)}

\newcommand\ie{{\it i.e.}~}

\usepackage{booktabs}
\usepackage{multirow}
\usepackage{dcolumn}
\usepackage{colortbl}
\newcommand\vertsp{\rule[-2mm]{1mm}{0mm} &}
\newcommand\horsp{\rule[-1.5mm]{0mm}{4.125mm}}
\newcommand\morehorsp{\rule[-2.25mm]{0mm}{6mm}}

\definecolor{magenta(process)}{rgb}{1.0, 0.0, 0.56}

\definecolor{darkspringgreen}{rgb}{0.09, 0.45, 0.27}

\definecolor{royalblue(web)}{rgb}{0.25, 0.41, 0.88}

\begin{document}

\title{Cosmic Microwave Background constraints on non-minimal couplings in inflationary models with power law potentials}

\author{Mehdi Shokri}
\email{mehdi.shokri@uniroma1.it}
\affiliation{Physics Department and INFN, Universit\`a di Roma ``La Sapienza'', Ple Aldo Moro 2, 00185, Rome, Italy}

\author{Fabrizio Renzi}
\email{fabrizio.renzi@roma1.infn.it}
\affiliation{Physics Department and INFN, Universit\`a di Roma ``La Sapienza'', Ple Aldo Moro 2, 00185, Rome, Italy}

\author{Alessandro Melchiorri}
\email{alessandro.melchiorri@roma1.infn.it}
\affiliation{Physics Department and INFN, Universit\`a di Roma ``La Sapienza'', Ple Aldo Moro 2, 00185, Rome, Italy}

\date{\today}

\preprint{}
\begin{abstract}
Inflationary models with power-law potentials are starting to be severely constrained by the recent measurements of Cosmic Microwave Background anisotropies provided by the Planck Satellite and by the BICEP2 telescope. In particular, models with power-law potentials $V(\varphi)\propto \varphi^n$ with $n \ge 2$ are strongly disfavored by present data since they predict a sizable contribution of gravitational waves with a tensor/scalar ratio of $r\sim0.15$ that is at odds with current limits. A non-minimal coupling to gravity has been proposed as a physical mechanism to lower the predictions for $r$. In this paper we further investigate the issue, presenting constraints on non-minimal couplings from current CMB data under the assumption of power-law potentials.

We found that models with $n>2$ show a statistically significant indication (above $95 \%$ C.L.) for a non minimal coupling. Non minimal coupling is also preferred by models with $n<2$ albeit just at about $68 \%$ C.L..
Interestingly, all the models considered show a non-zero running of the spectral index, $ n_{\rm run}$, consistent with the 2018 Planck release value of $-0.007 \pm 0.0068$.  We point out how future accurate measurement of $ n_{\rm run}$ would be necessary to significantly constraint these models and eventually rule out some or all of them. The combination of Planck data with the Bicep/Keck dataset strengthen these considerations. \\\\

{\bf PACS:} 98.80.-K; 98.80.Cq.
\\{\bf Keywords}: Non-minimal inflation, Jordan frame, Einstein frame, Power law potential, Observational constraints.
\end{abstract}

\maketitle

\section{Introduction}
The theory of cosmic inflation [1-7] provides a physical solution to several issues of the hot big bang cosmology such as the flatness, horizon and monopole problems by considering a period of accelerated expansion in the early universe. Also, it presents a viable mechanism to seed the primordial perturbations needed to form the observed large scale structures in our Universe (Cosmic Microwave Background anisotropies, galaxy clusters, filaments, etc.). Together with density fluctuations also a background of primordial gravitational waves (tensor perturbations) is produced (see e.g. [8-9]). Measuring such gravity waves background will represent an impressive confirmation of the inflationary scenario, and nevertheless, it will confirm the quantum nature of metric perturbations generated during inflation. Several experiments are currently being built aiming at their detection.

Several inflationary models have being conceived (see e.g. [10]).
According to a first classification of single field inflationary models, we can divide them into three main categories based upon the shape of potential: "large field", "small field" and "hybrid" models.

The large field models are based on a scalar field $\varphi$ with a power-law potential $V(\varphi)=c\varphi^{n}$ (see e.g. [11]). In these models, it is relatively easy to compute the predicted amount of primordial gravitational waves. The recent combined analysis of Planck 2018 and BICEP2/Keck Array Cosmic Microwave Background data [12] have indeed severely constrained these models, ruling out cases with $n\ge2$ that predict a too large GW signal, at odds with the current limits.
Indeed, for a $V(\varphi) =c \varphi^n$ model the tensor over scalar ratio $r$ can be approximately related to the spectral index of primordial perturbations $n_s$ (see e.g. [13]). Assuming the observed value of $n_s \sim 0.965$ and $n=4$
it is generally expected a GW contribution of $r\sim 0.20$, already in tension with the current observed limit of $r < 0.064$ at $95 \%$ C.L. [12].

Given their simplicity, several extensions to the $V(\varphi) = c\varphi^n$  models have been proposed to put them in better agreement with the current observational limits. For example, it has been shown that the inclusion of additional fields can reduce the predicted value of $r$ for $n=2$ models (see e.g. [14-16]). Another possibility is to consider an inflaton sound speed smaller than the speed of light due to a nonstandard kinetic term in the Lagrangian that describes the inflationary process [17].

In this work we take a different approach, investigating the possibility of a non-minimally coupling (NMC) term between the inflaton and the
Ricci scalar, i.e., between the gravity and matter sectors. In this scenario, already investigated by several authors (see e.g.[18-41]) the action is modified by adding a (NMC) term $\sim\frac{1}{2}\xi R\varphi^{2}$ where $R$ is the Ricci scalar and $\xi$ denotes the coupling constant.

While the introduction of this term complicates the simplicity of models based on power-law potentials, one has to bear in mind that the NMC term is unavoidable in some cases. For example, an NMC term arises naturally in the presence of quantum corrections in curved space-time, and it is necessary at the classical level for the renormalization of the theory [42]. More generally, as pointed out in [43], in any given inflationary scenario, the correct value of $\xi$ should be computed to have a theoretically consistent picture.

In this paper we provide constraints from CMB anisotropies data on the NMC term $\xi$ in the context of inflationary models with a power-law potential. We use as CMB data the Planck 2015 and the BICEP2/Keck array data releases ([44-45]).
The paper is structured as follows:
in the next Section, we derive, for several choices of the exponential $n$, the relations between $\xi$ and the number of e-foldings $N$, the two free parameters of the model, with the spectral index $n_S$, the running of the spectral index $\alpha_s$ and the tensor-to-scalar ratio $r$.
In Section III we discuss our analysis method, in Section IV we report our results and, finally, in Section V we present our conclusions.

\section{Minimal Coupling in power-law potentials}


 As stated in the introduction here we consider models characterized by the monomial potential $V(\varphi)=c\varphi^{n}$. The number n is usually a positive integer and here we assume the values $n=4,3,2,1$. However we also consider the interesting cases of  $n=2/3$ and $n=4/3$ that could arise, for example, in axion monodromy inflation (see e.g. [46-51]).
Let us consider the following action in the \textit{Jordan frame} (the main frame) for an inflationary model with a monomial potential and a NMC term:

\begin{equation}
S_J=\int{d^{4}x\sqrt{-g}\bigg(\frac{R}{2\kappa^{2}}-\frac{1}{2}g^{\mu\nu}\partial_{\mu}\varphi\partial_{\nu}\varphi-c\varphi^n+\frac{1}{2}\xi
R\varphi^{2}\bigg)}
\label{actionj1}
\end{equation}

\noindent where $\kappa^2=8\pi G$, $R$ is the Ricci scalar, and $\xi$ is the dimensionless coupling constant. In what follows, we are essentially interested in deriving a set of equations that could connect parameters as the scalar spectral index $n_s$, its running $\alpha_s$ and the tensor-to- scalar ratio $r$, that we can constrain by measuring, for example, the angular spectrum of the CMB anisotropies, to parameters of the inflationary model as $\xi$. To simplify the computations we derive the relevant equations in the \textit{Einstein frame} (the conformal frame) after the following conformal transformation:

\begin{equation}
\hat{g}_{\mu\nu}=\Omega^{2} g_{\mu\nu},
\end{equation}

with $\Omega^{2}=1+\kappa^{2}\xi\varphi^{2}$.
The action in the Einstein frame takes therefore the form:

\begin{equation}
S_{E}=\int
d^{4}x\sqrt{-\hat{g}}\bigg(\frac{\hat{R}}{2\kappa^{2}}-\frac{1}{2}F^{2}(\varphi)\hat{g}^{\mu\nu}\partial_{\mu}\varphi\partial_{\nu}\varphi
-\hat{V}(\hat{\varphi})\bigg).
\label{actione1}
\end{equation}

This conformal transformation is commonly used as a mathematical tool to map the equations of motion of physical systems into mathematically equivalent sets of equations that are more easily solved and computationally more convenient to study. Moreover, it has been entwined with gravitational theories (see e.g. [52-57]).
In the Einstein frame, the effective potential takes the form:

\begin{equation}
\hat{V}(\hat{\varphi})=\frac{c\varphi^{n}}{({1}+\kappa^{2}\xi\varphi^{2})^{2}}.
\end{equation}

\noindent Under the slow-roll approximation, the slow-roll parameters $\hat{\epsilon}$, $\hat{\eta}$,   and $\hat{\zeta}$ (see Eq.~\eqref{slowrollcondition} and Eq.~\eqref{slowroll} in the Appendix) can be written as

\begin{equation}
\hat{\epsilon}=\frac{\xi(n^{2}+2n(n-4)\kappa^{2}\xi\varphi^{2}+(n-4)^{2}\kappa^{4}\xi^{2}\varphi^{4})}{2\kappa^{2}\xi\varphi^{2}
({1}+\kappa^{2}\xi\varphi^{2}({1}+6\xi))},
\label{epsilon}
\end{equation}

\begin{align}
\hat{\eta}=\frac{1}{\kappa^{2}\xi\varphi^{2}(1+\kappa^{2}\xi\varphi^{2}(1+6\xi))^{2}}&\bigg[n(n-1)\xi+\xi\bigg(3n^{2}(1+2\xi)-2n(5+6\xi)-4\bigg)\kappa^{2}\xi\varphi^{2}+\label{eta}\\
+&\xi\bigg(3n^{2}(1+4\xi)-n(17+60\xi)+12\bigg)\kappa^{4}\xi^{2}\varphi^{4}+\xi\bigg((1+6\xi)(n^{2}-8n+16)\bigg)\kappa^{6}\xi^{3}\varphi^{6}\bigg] \notag
\end{align}

\begin{align}
\hat{\zeta}^{2}=\frac{1}{\kappa^{4}\varphi^{4}(1+\kappa^{2}\xi\varphi^{2}(1+6\xi))^{4}}&\bigg[\bigg(n^{2}(n-1)(n-2)\bigg)+\bigg(6n^{4}(1+2\xi)-n^{3}(31+54\xi)+2n^{2}(11+18\xi)-8n\bigg)\kappa^{2}\xi\varphi^{2}+\nonumber\\
+&\bigg(3n^{4}(12\xi^{2}+16\xi+5)-2n^{3}(108\xi^{2}+153\xi+55)+4n^{2}(72\xi^{2}+63\xi+41)+\nonumber\\
+&24n(1+4\xi)\bigg)\kappa^{4}\xi^{2}\varphi^{4}+\bigg(4n^{4}(36\xi^{2}+30\xi+5) - 2n^{3}(612\xi^{2}+546\xi+95) +\nonumber\\
+&4n^{2}(612\xi^{2}+612\xi+119)-4n(288\xi^{2}+120\xi+34)-32(7+12\xi)\bigg)\kappa^{6}\xi^{3}\varphi^{6}+\nonumber\\
+&\bigg(3n^{4}(72\xi^{2}+34\xi+5)-n^{3}(2376\xi^{2}+1152\xi+175)+2n^{2}(3744\xi^{2}+1944\xi+317)+\nonumber\\
-&4n(1296\xi^{2}+756\xi+158)-192(1+9\xi)\bigg)\kappa^{8}\xi^{4}\varphi^{8}+\bigg(6n^{4}(24\xi^{2}+10\xi+1)+\nonumber\\
-&n^{3}(1944\xi^{2}+822\xi+83)+n^{2}(8784\xi^{2}+3852\xi+398)-n(13248\xi^{2}+6528\xi+720)+\nonumber\\
+&288(1+6\xi)\bigg)\kappa^{10}\xi^{5}\varphi^{10}+\bigg((1+6\xi)^{2}(n^{4}-16n^{3}+96n^{2}-256n+256)\bigg)\kappa^{12}\xi^{6}\varphi^{12}\bigg]
\label{zeta}
\end{align}

Let us now consider the value of the inflaton field at the end of inflation $\varphi_e$ and at the start of inflation, i.e. at the time of Hubble Crossing, $\varphi_{HC}$.
Setting $\hat{\epsilon}=1$ at the end of inflation, and considering the quantity $\beta^{2}=\kappa^{2}\xi\varphi_{e}^{2}$, using Eq.~\ref{epsilon} we get:

\begin{equation}
\beta^{2}=\frac{\bigg(-(1-n(n-4)\xi)\pm\sqrt{1+8\xi
n+12\xi^{2}n^{2}}\bigg)}{(2(1+6\xi)-\xi(n-4)^{2})}.
\label{efolds}
\end{equation}

On the other hand, defining $m^{2}=\kappa^{2}\xi\varphi_{HC}^{2}$, we can write the slow-roll parameters at the beginning of inflation as

\begin{equation}
\hat{\epsilon}=\frac{\xi\(n+m^2(4 - n) \)^2}{2m^{2}({1}+m^{2}({1}+6\xi))}\ ,
\label{slwroll1}
\end{equation}
\begin{align}
\hat{\eta}=\frac{\xi}{m^{2}(1+m^{2}(1+6\xi))^{2}}
&\bigg[n(n-1)+\bigg(3n^{2}(1+2\xi)-2n(5+6\xi)-4\bigg)m^{2}+\nonumber\\
+&\bigg(3n^{2}(1+4\xi)-n(17+60\xi)+12\bigg)m^{4}+\bigg((1+6\xi)(n^{2}-8n+16)\bigg)m^{6}\bigg]\ ,
\end{align}
\begin{align}
\hat{\zeta}^{2}=\frac{\xi^{2}}{m^{4}(1+m^{2}(1+6\xi))^{4}}&\bigg[\bigg(n^{2}(n-1)(n-2)\bigg)+\bigg(6n^{4}(1+2\xi)-n^{3}(31+54\xi)+2n^{2}(11+18\xi)-8n\bigg)m^{2}+\nonumber\\
+&\bigg(3n^{4}(12\xi^{2}+16\xi+5)-2n^{3}(108\xi^{2}+153\xi+55)+4n^{2}(72\xi^{2}+63\xi+41)+24n(1+4\xi)\bigg)m^{4}+\nonumber\\
+&\bigg(4n^{4}(36\xi^{2}+30\xi+5)-2n^{3}(612\xi^{2}+546\xi+95)+4n^{2}(612\xi^{2}+612\xi+119)+\nonumber\\
-&4n(288\xi^{2}+120\xi+34)-32(7+12\xi)\bigg)m^{6}+\bigg(3n^{4}(72\xi^{2}+34\xi+5)-n^{3}(2376\xi^{2}+1152\xi+ \nonumber\\
+&175)+2n^{2}(3744\xi^{2}+1944\xi+317)-4n(1296\xi^{2}+756\xi+158)-192(1+9\xi)\bigg)m^{8}+\nonumber\\
+&\bigg(6n^{4}(24\xi^{2}+10\xi+1)-n^{3}(1944\xi^{2}+822\xi+83)+n^{2}(8784\xi^{2}+3852\xi+398)+\nonumber\\
-&n(13248\xi^{2}+6528\xi+720)+288(1+6\xi)\bigg)m^{10}+\bigg((1+6\xi)^{2}\times\notag \\
\times&(n^{4}-16n^{3}+96n^{2}-256n+256)\bigg)m^{12}\bigg].
\label{slwroll2}
\end{align}
The amount of inflation is usually specified considering the number of e-folds $ N $ defined as the logarithm of the ratio of the value of the scale factor at the end and beginning of inflation, \ie

\begin{equation}
e^{N}\equiv\frac{\hat{a}(\hat{t}_{e})}{\hat{a}(\hat{t}_{HC})}=\frac{a(t_{e})}{a(t_{HC})}\frac{\Omega(x_{end})}{\Omega(x_{HC})}
\label{efoldsone}
\end{equation}
where the hat denotes, as usual, the Einstein frame. It is well-know that the number of e-folds is strongly connected with the amount of perturbations generated during inflation and therefore to the cosmological parameters describing them. In NMC theories however also the coupling constant $ \xi $ enters those definition.
In the following, we will consider several value of the exponential $ n $ for the potential $ V(\phi) $, for each of them we will derive the relations connecting the coupling constant $ \xi $ and the number of e-folds $ N $ to the cosmological parameters $ r $, $ n_s $ and $ n_{\rm run} $. We will then use these relations to obtain constraints on the parameter space of $ \xi $ and $ N $ using Planck data.


\subsection{Case of $n=4$}

Probably the most famous form of large field inflationary potential is $V(\varphi)=\frac{1}{4}\lambda\varphi^{4}$. It corresponds to the quartic potential where the inflaton has a self-interacting feature. It is assumed that $\lambda < 1$ because otherwise, the interaction would become so strong that $\varphi$ would not correspond to a physical particle (the non-perturbative regime). On the other hand, values of $\lambda$ much smaller than $1$ are not usually envisaged since they would represent fine-tuning. The slow-roll parameters for this potential in the Einstein frame are given by Eqs.(\ref{epsilon} -- \ref{zeta}):

\begin{gather}
\hat{\epsilon}=\frac{8\xi}{m^{2}({1}+m^{2}({1}+6\xi))},\quad\quad \hat{\eta}=\frac{4\xi\bigg(3+m^{2}(1+12\xi)-2m^{4}(1+6\xi)\bigg)}{m^2(1+m^{2}(1+6\xi))^{2}},\label{epsilon-eta}
\\ \notag \\
\hat{\zeta}^{2}=\frac{32\xi^{2}\bigg(3+2m^{2}(-2+3\xi)-15m^{4}(1+6\xi)-6m^{6}(1+6\xi)^{2}+2m^{8}(1+6\xi)^{2}\bigg)}
{m^4({1}+m^{2}({1}+6\xi))^{4}}.\label{zeta2}
\end{gather}

In order to connected the number of e-folds with the inflaton field and the slow-roll parameters we need an expression for the scale factor $ a(t) $ during inflation. This can be found solving Eq.\eqref{friedmannI} under the slow-roll conditions, Eqs.\eqref{slowrollconditionham}, which left us with:
\begin{equation}
	\frac{a(t)}{a_{0}}=\(\frac{1+\kappa^{2}\xi\varphi^{2}(t)}{1+\kappa^{2}\xi\varphi^{2}_{0}}\)^{5/4}\exp\(\(\frac{1+6\xi}{8}\)\kappa^{2}\(\varphi_{0}^{2}-\varphi^{2}(t)\)\)
\end{equation}

\noindent where the subscribe "$0$" denotes the value of the inflaton field and scale factor at some time $ t_0 $. Taking $ t_0 $ to be the time of the Hubble crossing and using Eq.\eqref{efoldsone}, one obtains:
\begin{equation}\label{n4_efolds}
e^{N}=\(\frac{1+\beta^{2}}{1+m^{2}}\)^{5/4}\exp\({\frac{1+6\xi}{8\xi}\(m^{2}-\beta^{2}\)}\).
\end{equation}
for the e-folds number. If we now impose the consistency condition for large-potentials field $ m\geq\beta $, we find the relation:
\begin{equation}
	m^2 = \beta^2 + \frac{8\xi N}{1+6\xi}
\end{equation}
%


In what follows we restrict our analysis on the effect of non-minimal coupling under the approximation $ |\xi| \ll 1 $ and $ \psi \ll 1 $ \ie $ m^2 \ll 1 $. In this case the slow-roll parameters rewrite:
\begin{equation}
\hat{\epsilon}\simeq\frac{8\xi}{m^{2}},\quad\quad \hat{\eta}\simeq\frac{4\xi\bigg(3+m^{2}(1+12\xi)-2m^{4}(1+6\xi)\bigg)}{m^{2}}
\end{equation}
and
\begin{eqnarray}
\hat{\zeta}^{2}\simeq\frac{32\xi^{2}\bigg(3+2m^{2}(-2+3\xi)-15m^{4}(1+6\xi)-6m^{6}(1+6\xi)^{2}+2m^{8}(1+6\xi)^{2}\bigg)}
{m^{4}}.
\end{eqnarray}

\noindent We can now derive from the above equations the expressions for the scalar spectral index $n_s$, its running $\alpha_s = dn_s/d\log k$, and the tensor-to-scalar ratio $r$ such as:

\begin{subequations}
	\begin{equation}
	\hat{n}_{s}=1-6\hat{\epsilon}+2\hat{\eta}\simeq1-\frac{1}{N}(3-8\xi N),
	\label{eq.n4model_param1}
	\end{equation}
	
	\begin{equation}
	\hat{\alpha}_{s}=16\hat{\epsilon}\hat{\eta}-24\hat{\epsilon}^2-2\hat{\zeta}^{2}\simeq\frac{1}{N^{2}}(-3+96\xi N-64\xi^{2}N^{2}),
	\label{eq.n4model_param2}
	\end{equation}
	
	\begin{equation}
	\hat{r}=16\hat{\epsilon}
	\simeq\frac{16}{N}(1-8\xi N)
	\label{eq.n4model_param3}
	\end{equation}
\label{eq.n4model_param}	
\end{subequations}

The above equations can be reduced to $\hat{n}_{s}\simeq1-\frac{3}{N}$, $\hat{\alpha}_{s}\simeq-\frac{3}{N^{2}}$ and $\hat{r}\simeq\frac{16}{N}$ in the limit of $\xi\rightarrow 0$.

\subsection{Case of $n\neq4$ with $n\geq1$}

Following the same strategy, we continue here the analysis of the power-law potentials by considering other values of $n$. For potential with $ n \neq 4 $ we cannot use Eq.\eqref{n4_efolds}, therefore we need to restart by the definition of the e-folds number in th Einstein frame \ie

\begin{equation*}
N=-\sqrt{\frac{\kappa^{2}}{2}}\int\(\frac{1}{\sqrt{\hat{\epsilon}}}\)d\hat{\varphi},
\end{equation*}
which once integrated, gives :
\begin{equation*}
e^{N}=\(\frac{1+\beta^{2}}{1+m^{2}}\)^{\frac{5}{4}}\bigg(\frac{n+(n-4)\beta^{2}}{n+(n-4)m^{2}}\bigg)^{\({\frac{(n-4)-n(1+6\xi)}{8\xi(n-4)}}\)}
\end{equation*}
where $ \beta^2 $ is defined by Eq.\eqref{efolds}. Assuming again the consistency condition $m\geq\beta$, we obtain:

%

\begin{equation}
\bigg(\frac{n+(n-4)\beta^{2}}{n+(n-4)m^{2}}\bigg)\simeq e^{-2\xi N(n-4)} \,\,\ for \quad n\neq4.
\label{startofinflation}
\end{equation}

With this equation we can specify the expressions for the scalar spectral index $n_s$, its running $\alpha_s = dn_s/d\log k$, and the tensor-to-scalar ratio $r$ for each of the potential we are considering in the present work with $ n \neq 4 $.

\subsubsection{Case of $V\propto\varphi$}

%
%
%
%

In the case of $ n = 1  $ and with the assumption $ |\xi| \ll 1 $, we have from Eq.\eqref{startofinflation}:
\begin{equation*}
	m^{2}\simeq\frac{1}{3}\(1-(1-3\beta^{2})e^{-6\xi N}\) \simeq 2N\xi
\end{equation*}
%
%
For the spectral index, its running and the tensor-to-scalar ratio, using Eq.(\ref{epsilon-eta}) and Eq.\eqref{zeta2}, we have:
\begin{equation}\label{eq.n1_param}
\hat{n}_{s}\simeq1-\frac{1}{2N}(3+8\xi N),\quad \quad\hat{\alpha}_{s}\simeq\frac{1}{2N^{2}}(-3+6\xi N+4\xi^{2}N^{2}),\quad \quad\hat{r}\simeq\frac{4}{N}(1-12\xi N).
\end{equation}
Also, for $\xi\rightarrow 0$ the above equations are reduced to $\hat{n}_{s}\simeq1-\frac{3}{2N}$, $\hat{\alpha}_{s}\simeq-\frac{3}{2N^{2}}$ and $\hat{r}\simeq\frac{4}{N}$.

\subsubsection{Case of $V\propto\varphi^{2}$}

The simplest form of chaotic inflation is a non-interacting (free) field with a potential $V=\frac{1}{2}\mu^{2}\varphi^{2}$ where $\mu$ is the mass of the inflaton. The field equations have a time-independent, spatially homogeneous, solution $\varphi=0$, which represents the vacuum. Plane waves, related to oscillations around the vacuum state, correspond after quantization to non-interacting particles $\varphi$, which have mass $\mu$. Another feature of this potential is that in the presence of NMC between gravity and inflaton, the mass can be deformed to an effective mass by the shape of NMC term it is consequently more difficult to achieve slow-roll inflation.


%
%
%
%
\noindent In the case of $ n = 2  $ and with the assumption $ |\xi| \ll 1 $, we have from Eq.\eqref{startofinflation}:
\begin{equation*}
m^{2}\simeq 1-(1-\beta^{2})e^{-4\xi N} \simeq 4\xi N
\end{equation*}
The scalar spectral index, its running and tensor-to-scalar ratio are:
\begin{equation}\label{eq.n2_param}
\hat{n}_{s}\simeq1-\frac{2}{N}(1+\frac{4}{3}\xi^{2} N^{2}),\quad\quad \hat{\alpha}_{s}\simeq\frac{2}{N^{2}}(-1+4\xi N-96\xi^{2}N^{2}),\quad \quad\hat{r}\simeq\frac{8}{N}(1-8\xi N).
\end{equation}
and $\hat{n}_{s}\simeq1-\frac{2}{N}$, $\hat{\alpha}_{s}\simeq-\frac{2}{N^{2}}$ and $\hat{r}\simeq\frac{8}{N}$ when $\xi\rightarrow 0$.

\subsubsection{Case of $V\propto\varphi^{3}$}

%
%
\noindent In the case of $ n = 3  $ and with the assumption $ |\xi| \ll 1 $ we have from Eq.\eqref{startofinflation}:
\begin{equation*}
m^{2}=3-(3-\beta^{2})e^{-2\xi N} \simeq 6\xi N
\end{equation*}
In this case, the inflationary parameters take the following form
\begin{equation}\label{eq.n3_param}
\hat{n}_{s}\simeq1-\frac{1}{2N}(5-8\xi N),\quad\quad\hat{\alpha}_{s}\simeq\frac{5}{6N^{2}}(-3+42\xi N-468\xi^{2}N^{2}),\quad\quad\hat{r}\simeq\frac{12}{N}(1-4\xi N)
\end{equation}
For $\xi\rightarrow 0$, the above equations are expressed as $\hat{n}_{s}\simeq1-\frac{5}{2N}$, $\hat{\alpha}_{s}\simeq-\frac{5}{2N^{2}}$, and $\hat{r}\simeq\frac{12}{N}$.

\subsubsection{Case of $V\propto\varphi^{\frac{2}{3}}$}
\noindent In the case of $ n = 2/3  $ and with the assumption $ |\xi| \ll 1 $ we have from Eq.\eqref{startofinflation}:
\begin{equation*}
m^{2}=\frac{1}{5}\(1 - (1-5\beta^{2})e^{-20\xi N/3}\) \simeq \frac{4}{3}\xi N
\end{equation*}
The first order of spectral index, its running and tensor-to-scalar ratio are therefore:
\begin{equation}\label{eq.n23_param}
\hat{n}_{s}\simeq1-\frac{4}{3N}(1+4\xi N),\quad\quad\hat{\alpha}_{s}\simeq\frac{4}{81N^{2}}(-27+84\xi N+464\xi^{2}N^{2}),\quad\quad\hat{r}\simeq\frac{8}{9N}(3-40\xi N).
\end{equation}
In the limit $\xi\rightarrow 0$, we have $\hat{n}_{s}\simeq1-\frac{4}{3N}$, $\hat{\alpha}_{s}\simeq-\frac{4}{3N^{2}}$ and $\hat{r}\simeq\frac{8}{3N}$.

\subsubsection{$V\propto\varphi^{\frac{4}{3}}$}

\noindent In the case of $ n = 4/3  $ and with the assumption $ |\xi| \ll 1 $ we have from Eq.\eqref{startofinflation}:
\begin{equation*}
m^{2}=\frac{1}{2}\(1 - (1-2\beta^{2})e^{-\frac{16}{3}\xi N}\) \simeq \frac{8}{3}\xi N
\end{equation*}
For the inflationary parameters, we have
\begin{equation}\label{eq.n43_param}
\hat{n}_{s}\simeq1-\frac{1}{3N}(5+8\xi N),\quad\quad\hat{\alpha}_{s}\simeq\frac{5}{81N^{2}}(-27+48\xi N-704\xi^{2}N^{2}),\quad\quad\hat{r}\simeq\frac{16}{9N}(3-32\xi N).
\end{equation}
The above equations for $\xi\rightarrow 0$ are turned to $\hat{n}_{s}\simeq1-\frac{5}{3N}$, $\hat{\alpha}_{s}\simeq-\frac{5}{3N^{2}}$ and $\hat{r}\simeq\frac{16}{3N}$.

\section{Analysis Method}

As stated in the introduction, in this paper we place limits on the value of the coupling $\xi$ under the assumption of a particular inflationary model based on monomial potentials.
In general, CMB constraints on inflationary parameters are performed by letting the parameters $n_S$, $r$ and $\alpha_S$ to vary freely and by then comparing the predictions of a specific inflationary model with the allowed region of the parameters. Our approach here is different: an inflationary model is imposed {\it ab initio}, and we investigate the constraints on the parameters of that specific model. In particular, as we discussed in the previous section, our inflationary parameters are reduced to two: the number of e-foldings $N$ and the coupling term $\xi$.  While this kind of analysis is indeed more model dependent, it may provide constraints that are not achievable in a more general study where any value of $n_S$, $r$ and $\alpha_S$ is permitted.
Given a likelihood that compare data with theory\footnote{The theoretical models are computed using the latest version of the Boltzmann integrator CAMB [58].} constraints on cosmological parameters are extracted using the publicly available version of the Monte Carlo Markov Chain (MCMC) code CosmoMC [59] (Nov2016 version\footnote{https://cosmologist.info/}), based on the Metropolis-Hastings algorithm with chains convergence tested by the Gelman and Rubin method. We compare our theoretical models with data using the 2015 Planck likelihood, containing temperature and polarization spectra and their cross-correlation. We consider two cases for the Planck data: In the Planck high-$\ell$ case we consider only the CMB data at high multipoles $\ell>30$ and we impose an external prior on the optical depth $\tau=0.055\pm0.02$, i.e., we remove the large scale temperature and polarization data. In the Planck TTTEEE case, we consider the full Planck 2015 temperature and polarization dataset, including also the low multipoles and we disregard the prior on $\tau$.
Eventually, those two datasets are combined with the Bicep-Keck-Planck (BKP) B-mode likelihood [60].
We modified the code CosmoMC to accommodate $ N $ and $ \xi $ as independent parameters \ie they are randomly sampled in a given range, and to calculate the, now, derived parameters as function of the inflationary ones throughout Eq.(\ref{eq.n4model_param}) and Eqs.(\ref{eq.n1_param} -- \ref{eq.n43_param}). Note that in the publicly available version of CosmoMC the parameters $r$, $n_s$ and $\alpha_s$ are independent. An hard prior is imposed on the tensor-to-scalar ratio to assure is positiveness since for $ N\xi > \alpha^{-1}$, where $ \alpha $ is a constant value depending on the model we are considering, $ r $ is a negative quantity. The spectral index of tensor perturbations instead is evaluated using the inflationary consistency condition \ie $n_t = -r/8$ as in the standard version of CosmoMC. Along with the number of e-folds $ N $ and the coupling constant $ \xi $, we allow to vary the baryon $ \omega_b = \Omega_bh^2 $ and the CDM density $ \omega_c = \Omega_ch^2 $, the angular size of the sound horizon at decoupling $ \theta_s $, the reionization optical depth $\tau $ and the amplitude $ A_s $ and the spectral index $ n_s $ of scalar perturbations.
The assumed flat priors on these parameters are reported in Tab.\ref{tab:ranges}.

\begin{table*}[!hbtp]
	\begin{center}
	   \begin{tabular}{cc}
			\toprule
			\horsp
			Parameter \vertsp Prior\\
			\hline
			\hline
			\morehorsp
			$ \omega_b $ \vertsp $ [0.02 \div 0.25] $\\
			$ \omega_c $ \vertsp $ [0.1 \div 0.3] $ \\
			$ \theta_s $ \vertsp $ [0.5 \div 2] $ \\		        			$ \tau $ \vertsp $ [0.01\div 0.8] $ \\
			$ \ln(10^{10}A_s) $ \vertsp $ [3.01 \div 3.2] $ \\
			$ N $ \vertsp $ [20 \div 100] $ \\
	     	$ \xi $ \vertsp $ [-0.2 \div 0.2] $ \\
			\bottomrule
 	  \end{tabular}
	\end{center}
    \caption{Range of the flat prior on the parameters varied in the MCMC analysis.}		\label{tab:ranges}	
\end{table*}	

\section{Results}

The constraints on the inflationary parameters from the Planck 2015 data and from their combination with the BICEP2/Keck Array release are reported in Table~\ref{Table1}. In Figure~\ref{fig1a} and Figure~\ref{fig1} we show the contour plots at $68 \%$ and $95 \%$ C.L. from the Planck high$\ell$ and Planck TTTEEE data, respectively. In Figure~\ref{fig2a} and Figure~\ref{fig2} we show the analogous constraints obtained now with the inclusion of the BKP dataset.

\begin{figure*}[!hbtp]
	\centering
	\begin{subfigure}{.495\textwidth}
		\includegraphics[width=\textwidth,keepaspectratio]{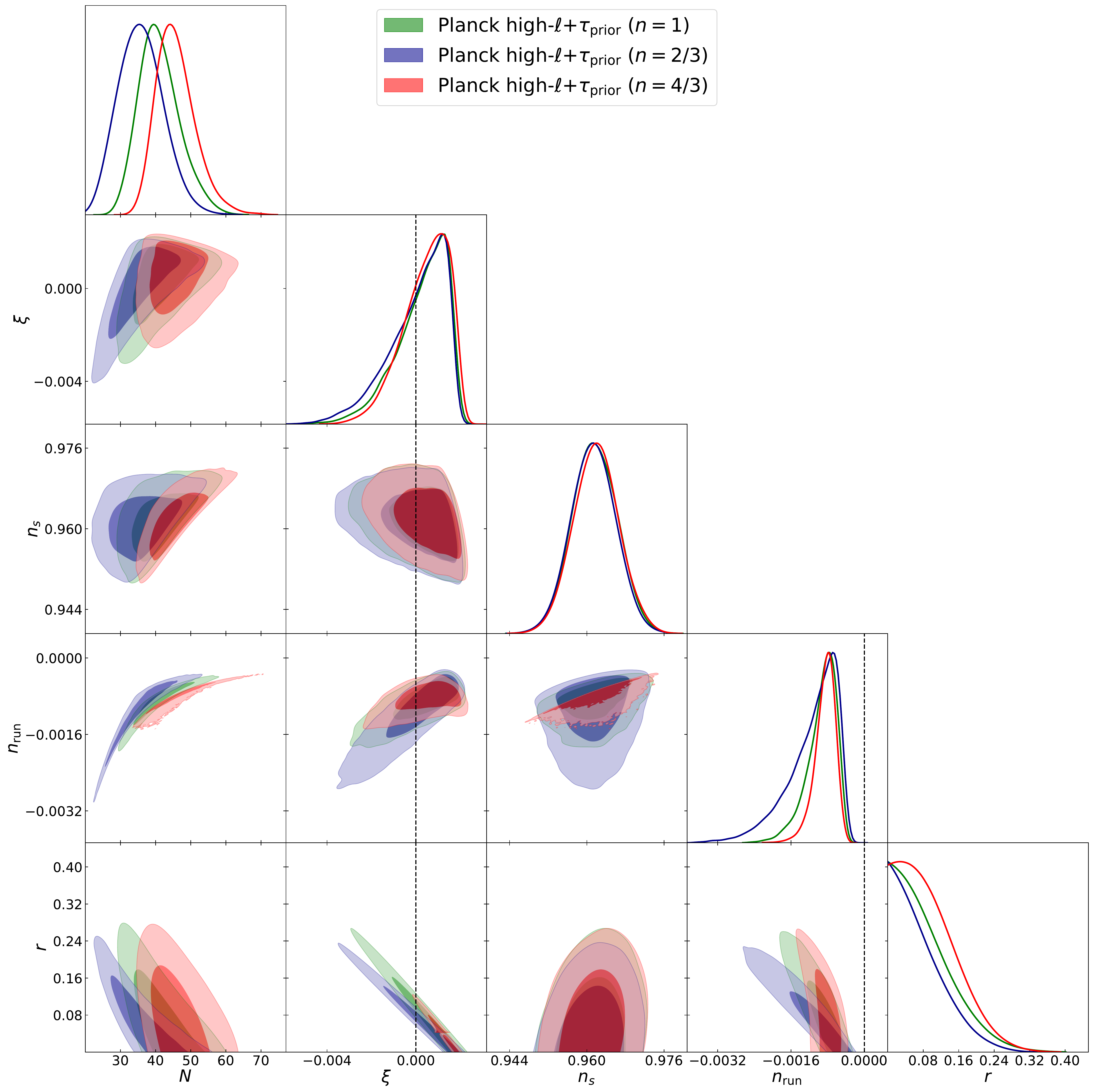}
	\end{subfigure}
	\begin{subfigure}{0.495\textwidth}
	    \includegraphics[width=\textwidth,keepaspectratio]{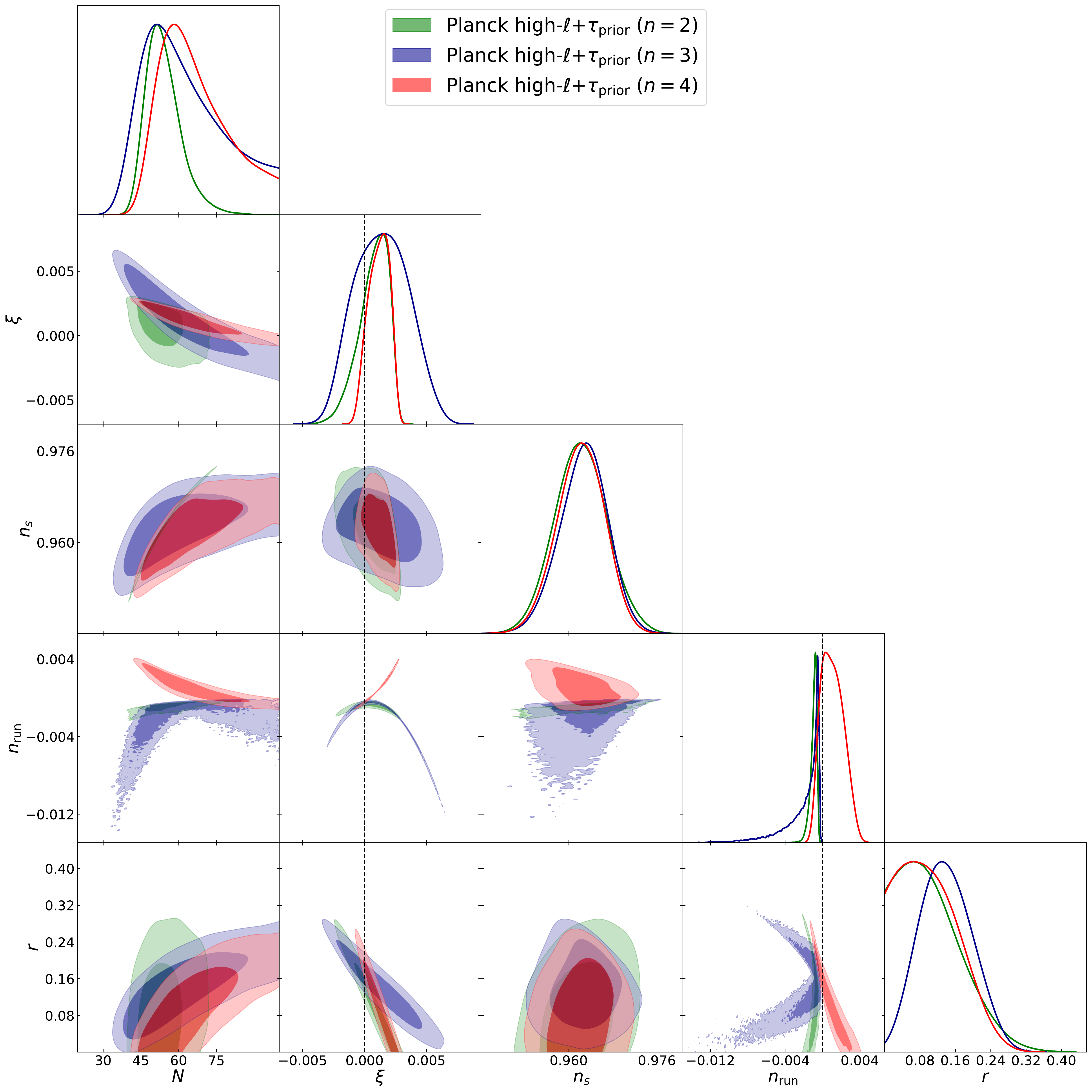}
	\end{subfigure}

	\caption{Constraints at $68 \%$ and $95 \%$ C.L. on cosmological and inflationary parameters from Planck anisotropy and polarization data at high multipoles ($\textit{l}>30$) with the inclusion of a prior on the reionization optical depth. Power-law potentials with $n < 2$ are on the left and power-law potentials with $n \ge 2$ are on the rigth.}
	\label{fig1}
\end{figure*}

\begin{figure*}[!hbtp]
	\centering
	\begin{subfigure}{.495\textwidth}
		\includegraphics[width=\textwidth,keepaspectratio]{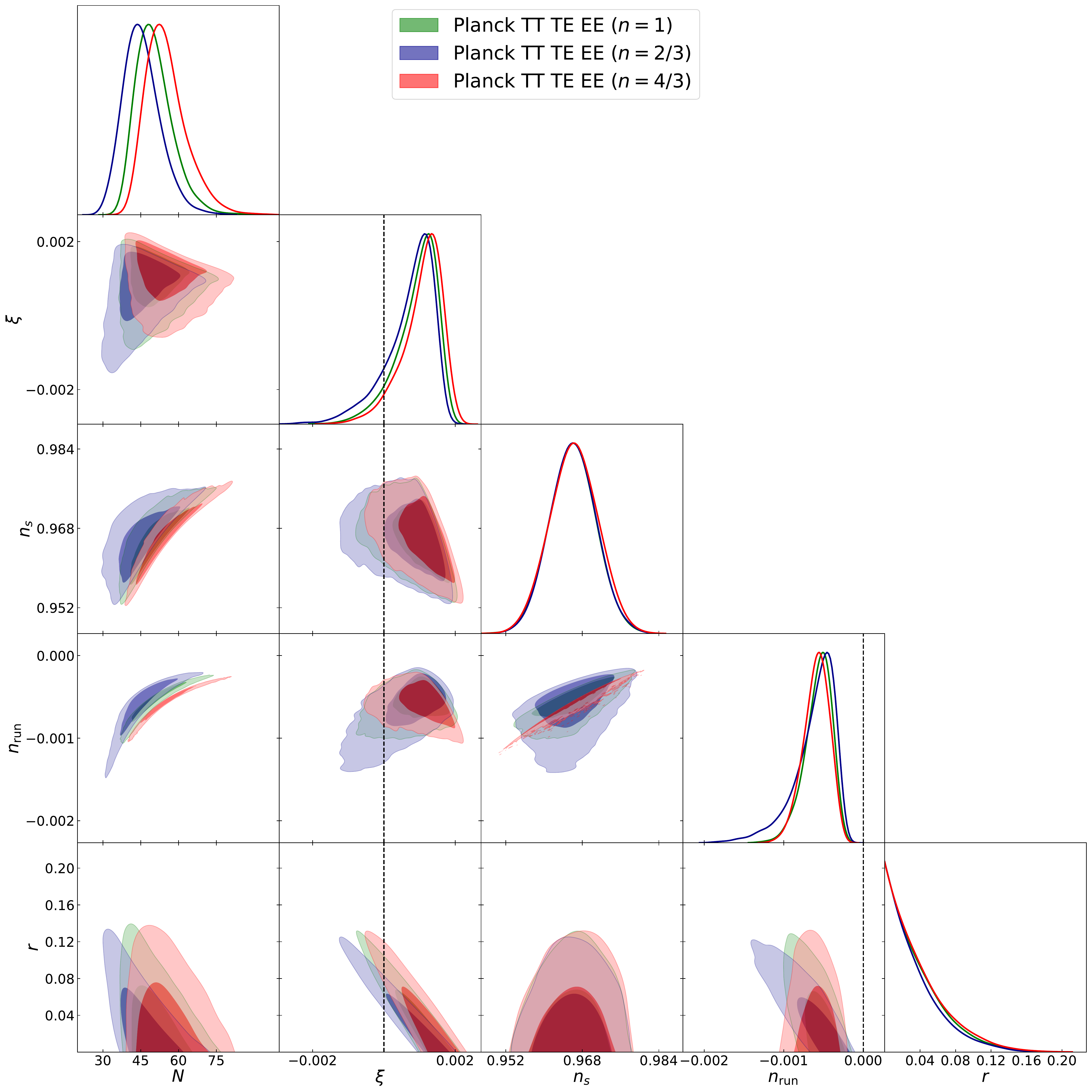}
	\end{subfigure}
	\begin{subfigure}{0.495\textwidth}
	    \includegraphics[width=\textwidth,keepaspectratio]{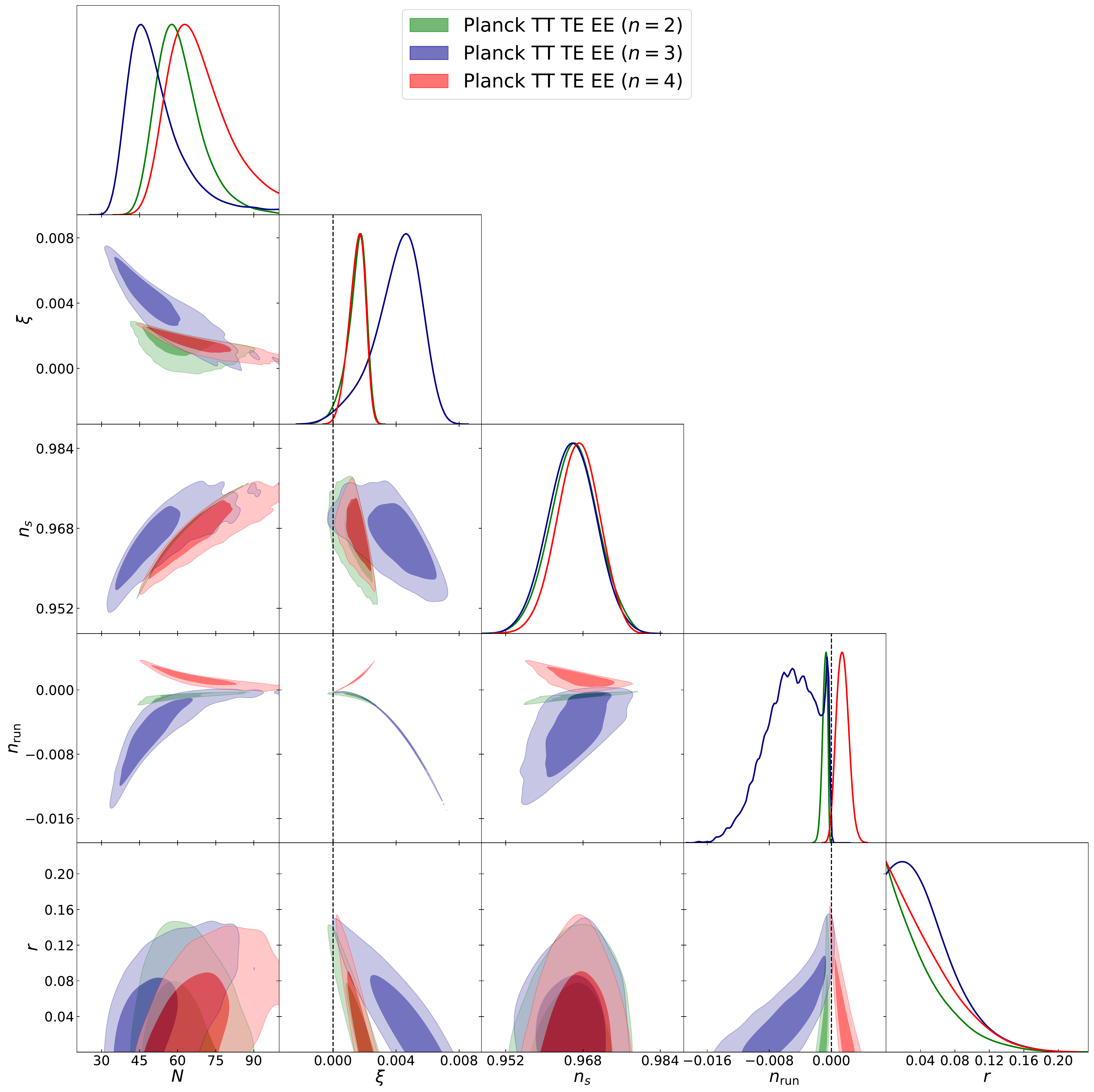}
	\end{subfigure}

	\caption{Constraints at $68 \%$ and $95 \%$ C.L. on cosmological and inflationary parameters from Planck anisotropy and polarization data. Power-law potentials with $n < 2$ are on the left and power-law potentials with $n \ge 2$ are on the rigth.}
	\label{fig1}
\end{figure*}

\begin{figure*}[!hbtp]
	\centering
	\begin{subfigure}{.49\textwidth}
		\includegraphics[width=\textwidth,keepaspectratio]{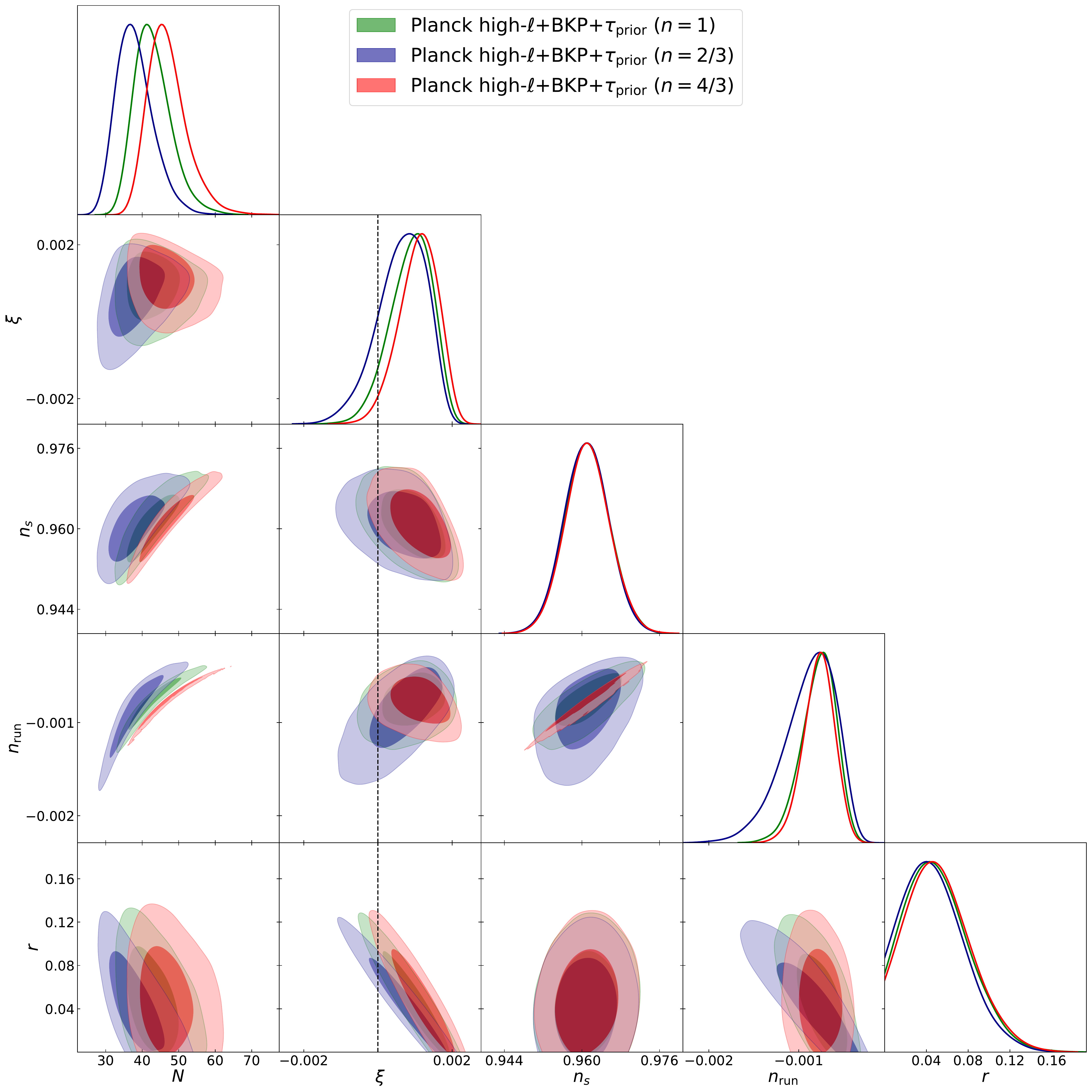}
	\end{subfigure}
	\begin{subfigure}{0.49\textwidth}
		\includegraphics[width=\textwidth,keepaspectratio]{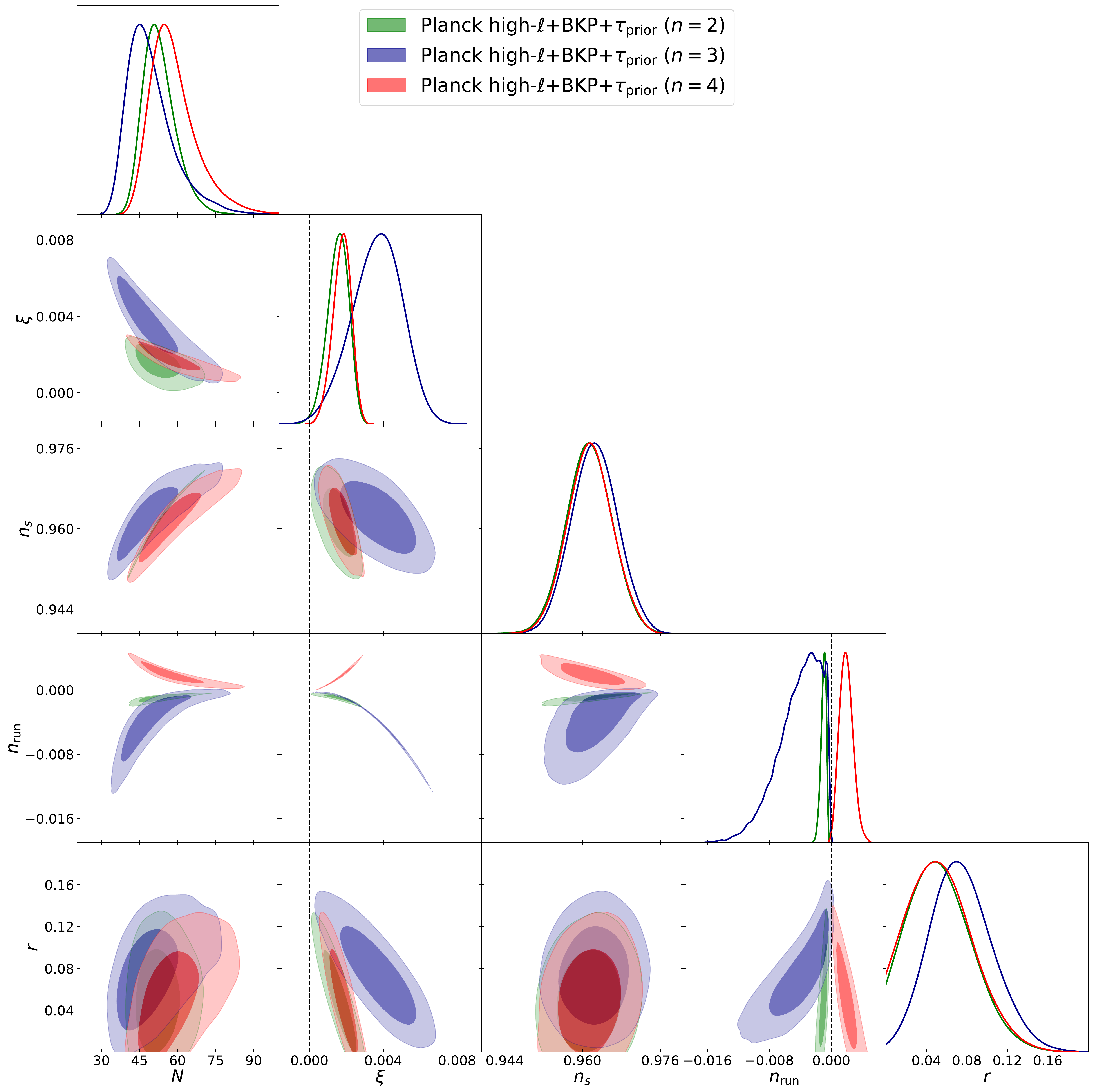}
	\end{subfigure}
	
	\caption{Constraints at $68 \%$ and $95 \%$ C.L. on cosmological and inflationary parameters from Planck anisotropy and polarization data combined with the BKP likelihood and with the inclusion of a prior on the reionization optical depth. Power-law potentials with $n < 2$ are on the left and power-law potentials with $n \ge 2$ are on the rigth.}
	\label{fig2a}
\end{figure*}

\begin{figure*}[!hbtp]
	\centering
	\begin{subfigure}{.49\textwidth}
		\includegraphics[width=\textwidth,keepaspectratio]{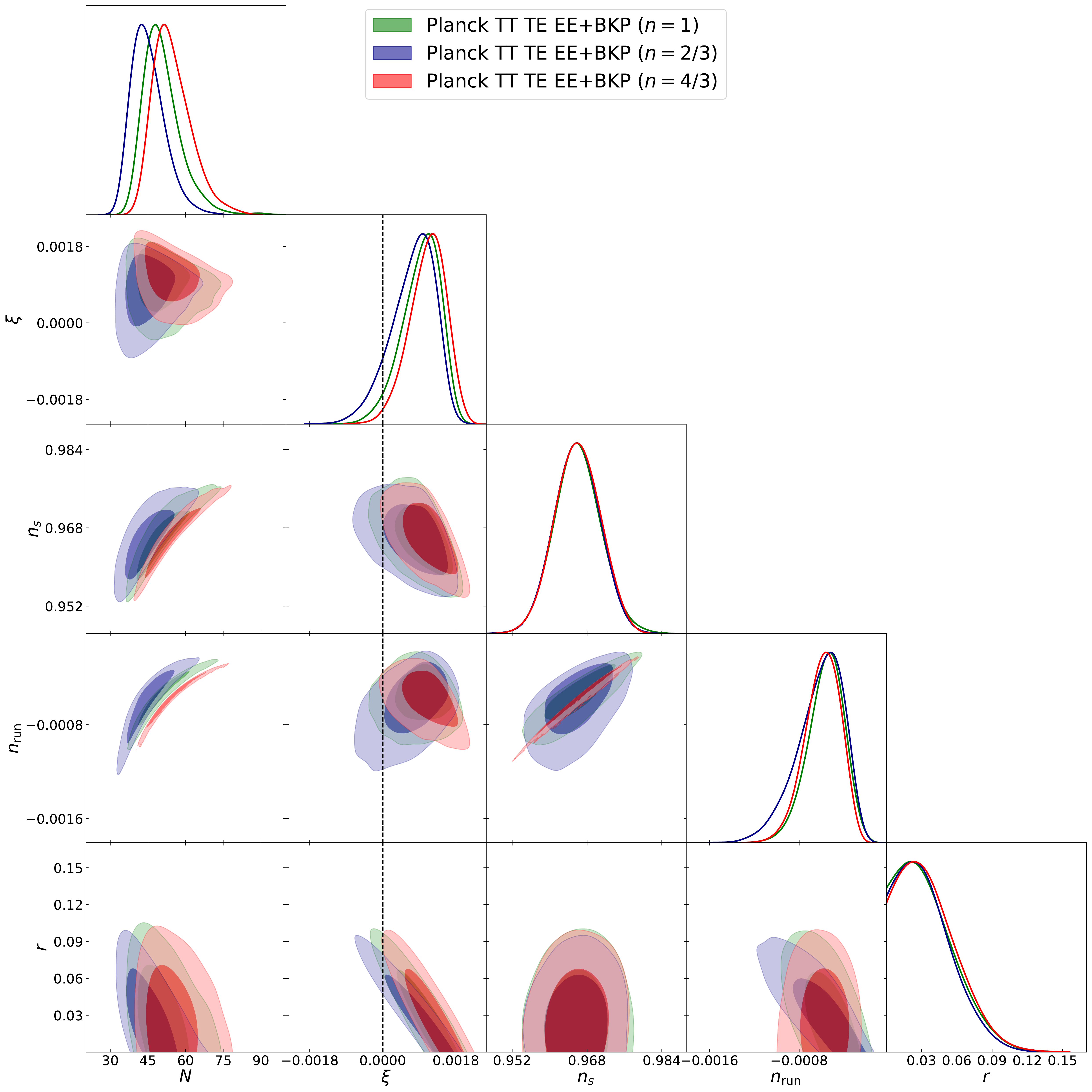}
	\end{subfigure}
	\begin{subfigure}{0.49\textwidth}
		\includegraphics[width=\textwidth,keepaspectratio]{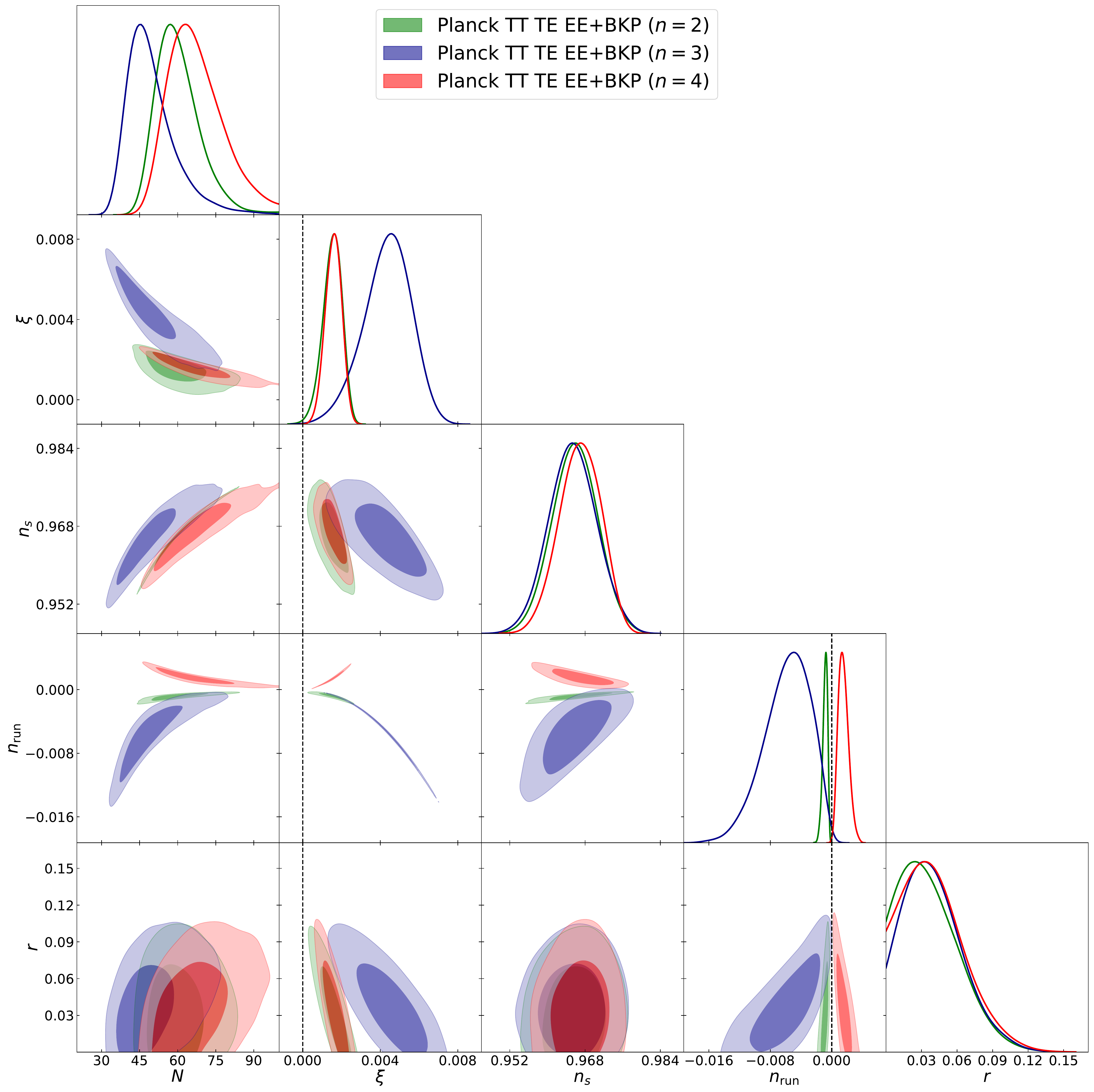}
	\end{subfigure}
	
	\caption{Constraints at $68 \%$ and $95 \%$ C.L. on cosmological and inflationary parameters from Planck anisotropy and polarization data combined with the BKP likelihood. Power-law potentials with $n < 2$ are on the left and power-law potentials with $n \ge 2$ are on the rigth.}
	\label{fig2}
\end{figure*}

\begin{table*}[!hbtp]
	\begin{center}
		\begin{tabular}{ccccccc}
			\toprule
			\vertsp Parameter \vertsp Planck high-$\ell$+$\tau_{\rm prior}$ \vertsp Planck TTTEEE \vertsp Planck high-$\ell$+$\tau_{\rm prior}$+BKP \vertsp Planck TTTEEE+BKP \\
			\hline\hline
			\morehorsp
			n = 1   \vertsp N \vertsp $ 41^{+5}_{-7} $ \vertsp $ 50^{+5}_{-9} $\vertsp $ 43^{+4}_{-6} $  \vertsp $ 51^{+5}_{-9} $ \\
			\vertsp $ \xi $ \vertsp $ 0.0003^{+0.0014}_{-0.0006} $ \vertsp $ 0.0009^{+0.0007}_{-0.0003} $ \vertsp $ 0.0009^{+0.0007}_{-0.0005} $ \vertsp $ 0.0009^{+0.0006}_{-0.0004} $ \\
			\vertsp $ n_s $ \vertsp $ 0.962\pm 0.005 $ \vertsp $ 0.966\pm 0.005 $ \vertsp $ 0.961\pm 0.005 $ \vertsp $ 0.966\pm 0.005$ \\
			\vertsp $ n_{\rm run} $ \vertsp $ -0.0009^{+0.0004}_{-0.0002} $ \vertsp $ -0.0006^{+0.0002}_{-0.0001} $ \vertsp $ -0.0008\pm{-0.0002} $ \vertsp $ -0.0006^{+0.0002}_{-0.0001} $\\	
			\vertsp $ r $ \vertsp $ < 0.213 $ \vertsp $ < 0.100 $ \vertsp $ < 0.106 $ \vertsp $ <0.081 $\\
            \vertsp $ \chi^2 $ \vertsp $ 2453 $ \vertsp $ 12992 $ \vertsp $ 2495 $ \vertsp $ 12992 $\\
			\hline\hline
			\morehorsp
			n = 2/3\vertsp N \vertsp $ 36^{+6}_{-7} $ \vertsp $ 45^{+6}_{-8} $ \vertsp $ 38^{+4}_{-6} $ \vertsp $ 45^{+5}_{-8} $ \\
			\vertsp $ \xi $ \vertsp $ 0.0000^{+0.0016}_{-0.0006} $ \vertsp $ 0.0008^{+0.0008}_{-0.0003} $ \vertsp $ 0.0006^{+0.0008}_{-0.0005} $ \vertsp $ 0.0007^{+0.0007}_{-0.0004}$ \\
			\vertsp $ n_s $ \vertsp $ 0.961\pm {0.005}$ \vertsp $ 0.966^{+0.005}_{-0.004} $ \vertsp $ 0.961\pm{0.005} $ \vertsp $ 0.966\pm 0.005$ \\
			\vertsp $ n_{\rm run} $\vertsp $ -0.0011^{+0.0007}_{-0.0002} $ \vertsp $ -0.0006^{+0.0003}_{-0.0001} $ \vertsp $ -0.0009^{+0.0004}_{-0.0002} $ \vertsp $ -0.0006^{+0.0003}_{-0.0001} $ \\	
            \vertsp $ r $ \vertsp $ < 0.188 $ \vertsp $ < 0.094 $  \vertsp $ <0.101 $ \vertsp $ <0.077 $\\
			\vertsp $ \chi^2 $ \vertsp $ 2452 $ \vertsp $ 12949 $ \vertsp $ 2495 $ \vertsp $ 12992 $\\
			\hline\hline
			\morehorsp
			n = 4/3\vertsp N \vertsp $ 46^{+4}_{-6} $ \vertsp $ 55^{+5}_{-10} $ \vertsp $ 47^{+4}_{-6} $ \vertsp $ 55^{+5}_{-9} $\\
			\vertsp $ \xi $ \vertsp $ 0.0004^{+0.0013}_{-0.0006} $ \vertsp $ 0.0011^{+0.0007}_{-0.0003} $ \vertsp $ 0.0011^{+0.0006}_{-0.0004} $ \vertsp $0.0011\pm{0.0005}$ \\
			\vertsp $ n_s $ \vertsp $ 0.962\pm 0.005$ \vertsp $ 0.966\pm 0.005 $ \vertsp $ 0.961\pm 0.005$ \vertsp $ 0.966\pm 0.005 $ \\
			\vertsp $ n_{\rm run} $ \vertsp $ -0.0009\pm{0.0003} $ \vertsp $ -0.0006^{+0.0002}_{-0.0001} $ \vertsp $ -0.0008\pm{0.0002} $ \vertsp $-0.0006^{+0.0002}_{-0.0001}$\\	
			\vertsp $ r $ \vertsp $ < 0.215 $ \vertsp $ < 0.106 $ \vertsp $ <0.108 $ \vertsp $ < 0.082 $ \\
            \vertsp $ \chi^2 $ \vertsp $ 2454 $  \vertsp $ 12950 $ \vertsp $ 2495 $ \vertsp $ 12992$\\
			\bottomrule
		\end{tabular}
        \end{center}
        \begin{center}
		\begin{tabular}{ccccccc}
			\toprule
			\vertsp Parameter \vertsp Planck high-$\ell$+$\tau_{\rm prior}$ \vertsp Planck TTTEEE \vertsp Planck high-$\ell$+$\tau_{\rm prior}$+BKP \vertsp Planck TTTEEE+BKP \\
			\hline\hline
			\morehorsp
			n = 2   \vertsp N \vertsp $ 54^{+5}_{-8} $ \vertsp $ 61^{+6}_{-11} $ \vertsp $ 53^{+5}_{-8} $ \vertsp $ 60^{+6}_{-10}$\\
			\vertsp $ \xi $ \vertsp $ 0.0007^{+0.0015}_{-0.0008} $ \vertsp $ 0.0014^{+0.0008}_{-0.0004} $ \vertsp $ 0.0015\pm{0.0006} $ \vertsp $ 0.0015\pm{0.0005}$ \\
			\vertsp $ n_s $ \vertsp $ 0.962\pm 0.005 $ \vertsp $ 0.966\pm 0.005 $ \vertsp $ 0.961\pm 0.005 $ \vertsp $ 0.966\pm 0.005$ \\
			\vertsp $ n_{\rm run} $ \vertsp $ -0.0010^{+0.0004}_{-0.0002} $ \vertsp $ -0.0009^{+0.0004}_{-0.0002} $ \vertsp $ -0.0010^{+0.0004}_{-0.0002} $ \vertsp $ -0.0009^{+0.0004}_{-0.0002}$\\	
			\vertsp $ r $ \vertsp $ < 0.238 $ \vertsp $ < 0.115 $ \vertsp $ <0.109 $ \vertsp $ <0.087 $\\				   		     	         \vertsp $ \chi^2 $ \vertsp $ 2452 $  \vertsp $ 12950 $ \vertsp $ 2495 $ \vertsp $ 12992 $\\
			\hline\hline
			\morehorsp
			n = 3 \vertsp N \vertsp $ 62^{+10}_{-21} $ \vertsp $ 52^{+5}_{-13} $ \vertsp $ 53^{+5}_{-8} $ \vertsp $ 51^{+5}_{-12}$\\
			\vertsp $ \xi $ \vertsp $ 0.0014\pm{0.0023} $ \vertsp $ 0.0040^{+0.0018}_{-0.0011} $ \vertsp $ 0.0036^{+0.0015}_{-0.0012} $ \vertsp $ 0.0042^{+0.0014}_{-0.0010}$ \\
			\vertsp $ n_s $ \vertsp $ 0.963^{+0.005}_{-0.004}$ \vertsp $ 0.966\pm{+0.005} $ \vertsp $ 0.962\pm{+0.005} $ \vertsp $ 0.966\pm{+0.005}$ \\
			\vertsp $ n_{\rm run} $ \vertsp $ -0.0022^{+0.0019}_{-0.0002} $ \vertsp $ -0.0053^{+0.0050}_{-0.0021} $ \vertsp $ -0.0042^{+0.0038}_{-0.0012}$ \vertsp $ -0.0055^{+0.0035}_{-0.0024}$ \\	
			\vertsp $ r $ \vertsp $ 0.139^{+0.056}_{-0.066} $ \vertsp $ < 0.123 $ \vertsp $ 0.075^{+0.027}_{-0.034} $ \vertsp $< 0.087 $\\		
            \vertsp $ \chi^2 $ \vertsp $ 2456 $ \vertsp $ 12950 $  \vertsp $ 2497 $ \vertsp $ 12991 $\\
			\hline\hline
			\morehorsp
			n = 4 \vertsp N \vertsp $ 66^{+9}_{-17} $ \vertsp $66^{+8}_{-14}$ \vertsp $ 59^{+6}_{-11} $ \vertsp $67^{+8}_{-13}$ \\
			\vertsp $ \xi $ \vertsp $ 0.0012^{+0.0010}_{-0.0014} $ \vertsp $0.0015^{+0.0006}_{-0.0004}$ \vertsp $ 0.0018\pm{+0.0005}$ \vertsp $0.0016\pm0.0004$ \\
			\vertsp $ n_s $ \vertsp $ 0.962^{+0.005}_{-0.004} $ \vertsp $0.966{\pm 0.004}$ \vertsp $ 0.962\pm 0.005$ \vertsp $0.967^{+0.005}_{-0.004}$ \\
			\vertsp $ n_{\rm run} $ \vertsp $ 0.0011^{+0.0010}_{-0.0014} $ \vertsp $0.0014^{+0.0007}_{-0.0008}$ \vertsp $ 0.0019^{+0.001}_{-0.001}$ \vertsp $0.0015^{+0.0005}_{-0.0008}$ \\	
			\vertsp $ r $ \vertsp $ <0.219 $ \vertsp $<0.121$ \vertsp $ <0.111 $ \vertsp $<0.089$ \\
            \vertsp $ \chi^2 $ \vertsp $ 2453 $ \vertsp $12950$ \vertsp $ 2495 $ \vertsp $12993$ \\
			\bottomrule
		\end{tabular}
	\end{center}
	\caption{Constraints on cosmological and inflationary parameters in case of power-law potentials with non-minimal coupling from Planck and Planck+BKP datasets. Constraints on parameters are at the $68 \% $ C.L.  (upper limits at $95 \%$ C.L.)  }
    \label{Table1}
\end{table*}

Let us first consider the results obtained from the Planck 2015 datasets (with and without the low multipoles data) alone.
As we can see from the first column of Table~\ref{Table1} and Figure~\ref{fig1a}, we found no evidence for a coupling ($\xi\neq0$) from the Planck high-$\ell$+$\tau_{prior}$ data in any of the power-law models considered. Moreover, by looking at the reported values of the $\chi^2_{eff}$, we see that models with $n>2$ have a $\Delta \chi^2 \sim 4$ with respect to models with $n=1$, i.e., they provide a worse fit to the data at about two standard deviations. In practice, the Planck high-$\ell$ data alone is unable to rule out significantly models with $n=2,3,4$. This fact is mainly due to the poor constraints on the tensor to scalar ratio $r$ achievable from this dataset. It is however compelling, that all models, except for the $n=2/3$ case, shows an indication for a negative running $n_{run}\sim -0.001$. This result is not due to an actual presence of running in the data but to the specific correlations between $n_{run}$ and the other inflationary parameters present in the models considered. So one should be careful in claiming any general indication for $n_{run}$ from this analysis. However, this shows either the potential of future measurements of $n_{run}$ of discriminating between these models, either the fact that a running at this level could be easily produced and that it should not be discarded in the analysis of future data.

As we can see from the second column of Table~\ref{Table1} and Figure~\ref{fig1}, the inclusion of the low multipole CMB data, without a prior on the optical depth, has the main effect of substantially increasing (by a factor $\sim 2$) the constraint on $r$. The main consequence of this is that in this case, an indication for a coupling $\xi$ starts to emerge. If we consider the values reported in Table~\ref{Table1} and the corresponding posteriors plotted in Figure~\ref{fig1} (left panel) we see that for models with $n<2$, the indication is slightly above one standard deviation (consider that the posterior on $\xi$ is non-gaussian in this case), while, considering the posteriors $\xi$ in Figure~\ref{fig2}, right panel, it is above the two standard deviations for $n> 2$
(and close to two standard deviation for $n=2$). Again, as we pointed out in the previous paragraph, this indication for $\xi\neq0$ is not generic and must be considered valid only for models with an NMC term and power-law potential with $n>2$. Considering the values of the $\chi^2_{eff}$ we see that they are almost identical for any value of $n$ considered. In few words, the inclusion of an NMC term at the level of $\xi\sim 0.004$ makes models with $n=2,3,4$ back into agreement with the full Planck 2015 dataset.
Considering the running, we can also notice that models with $n<2$ all show an indication for a negative running but at the level of $n_{run}\sim 0.0006$. Models with $n\ge2$ show on the contrary a significantly lower negative running with $n_{run}\sim0.001$. Again, a future accurate measurement of $n_{run}$ could significantly discriminate between inflationary models.

In the third and fourth columns of Table~\ref{Table1} we report the constraints obtained by combing the Planck 2015 data with the BKP dataset. As expected, the inclusion of the BKP dataset significantly increase the limits on $r$. It is interesting to notice that the constraint on $r$ from the full Planck dataset are similar to those obtained by the Planck high-$\ell$+BKP dataset, showing a good agreement between the low multipole data from Planck and BKP.
As we can see from the results reported in Table~\ref{Table1} and the posteriors in Figure~\ref{fig2a}, and Figure~\ref{fig2}, the inclusion of the BKP dataset improves the indication for $\xi>0$ obtained from the Planck dataset alone. We have now from the Planck high-$\ell$+BKP dataset an indication for coupling above one standard deviation for $n=1$ and $n=4/3$, at about two standard deviation for $n=2$, and, finally, above $95 \%$ C.L. for $n=3$ and $n=4$. When the Planck TTTEEE+BKP dataset is considered, the indication for $\xi$ is above one standard deviation for $n=1$ and $n=2/3$, at about two standard deviations for $n=4/3$ and above two standard deviations for $n\ge2$
Considering now the constraints on the running of the spectral index $n_{run}$, we see that while models with $n\le 2$ prefer a running around $n_{run}\sim -0.001$ at the $95 \%$ C.L., models with $n>2$ are suggesting an higher value around $n_{run}\sim - 0.006$. These values are both consistent with the latest constraints from Planck ($n_{run}=-0.007\pm0.0068$, see [12]) and clearly indicates that future constraints on $n_{run}$ could significantly constrain models with NMC.

\section{Conclusions}

In this paper, we have compared with the recent observations a particular class of inflationary models: i.e., models with a power-law potential and an NMC. The primary motivation for studying these scenarios is that inflationary models with a power-potential generally predict a too large amount of gravitational waves to be consistent with current limits, while the inclusion of an NMC can in principle put them back into agreement with observations.
Our results can be summarized as follows:

\begin{itemize}

\item If we conservatively consider only the Planck data at high-$\ell$ plus an external prior on the optical depth, the bounds on $r$ are rather weak, and we found no indication for coupling from this dataset.

\item If consider the full Planck dataset we obtain an indication for a coupling $\xi \sim 0.001$ at the level above one standard deviation for power-law potentials with $n=1,2/3,4/3$, and a sign for a more substantial coupling in the range $\xi \sim 0.002-0.004$ for $n=2,3,4$ at two standard deviations. These results are confirmed and reinforced by the inclusion of the BKP data.

\item The models considered also show a significant running $n_{run}$. When we consider the full Planck dataset in combination with the BKP dataset we get an indication above two standard deviations for running $n_{run}\sim -0.0006$ for $n<2$ and for larger negative running in the range $[-0.007\, ;\, -0.001]$ from $n\ge2$.

\end{itemize}

In this paper, we have therefore not only confirmed that NMC inflationary models with a power law potential with $n \ge 2$ could provide a good fit to current Planck+BKP data but also obtained constraints on the value of the coupling $\xi$ needed to achieve this result. Moreover, we have found that models with $n< 2$ predict a negative value of the running of the spectral index of $n_{run} \sim -0.0006$ while models with $n\ge2$ predict a even more negative value in the range  $n_{run}\sim -0.0015\, :\,-0.006$. Given the current constraints from Planck on $n_{run}$ that show a sensitivity of  $\Delta n_{run}\sim0.007$ is therefore possible that near future measurements could significantly constrain power law NMC models with $n\ge 2$.

\acknowledgments

AM thanks the University of Manchester and the Jodrell Bank Center for Astrophysics for hospitality. AM and FR are supported by TASP, iniziativa specifica INFN. We thank Flavio Bombacigno, Fabio Moretti and Matteo Becchetti for useful comments.

\appendix
\section{The equivalence between Jordan and Einstein frames}
\subsection{Jordan Frame}

Let us consider the following action
in the Jordan frame for a generic inflationary model with a NMC term:

\begin{equation}
S_J=\int{d^{4}x\sqrt{-g}\bigg(\frac{R}{2\kappa^{2}}-\frac{1}{2}g^{\mu\nu}\partial_{\mu}\varphi\partial_{\nu}\varphi-V(\varphi)+\frac{1}{2}\xi
R\varphi^{2}\bigg)}
\label{actionj}
\end{equation}

\noindent where $\kappa^2=8\pi G$, $R$ is the Ricci scalar, and $\xi$ is the dimensionless coupling constant. A variation of the action with respect to the scalar field $\varphi$, we obtain the following evolution equation for the inflaton field:
\begin{eqnarray}
\ddot{\varphi}+3H\dot{\varphi}+\bigg(\frac{\kappa^{2}\xi\varphi^{2}(1+6\xi)}
{1+\kappa^{2}\xi\varphi^{2}(1+6\xi)}\bigg)\frac{\dot{\varphi}^{2}}{\varphi}=\bigg(\frac{4\kappa^{2}\xi\varphi
V(\varphi)-(1+\kappa^{2}\xi\varphi^{2})\frac{dV}{d\varphi}}{1+\kappa^{2}\xi\varphi^{2}({1}+6\xi)}\bigg)
\label{kleingordon}
\end{eqnarray}
where overdots denote derivatives with respect to the coordinate time. Instead, varying the action $S_J$ with respect to the metric $g_{\mu\nu}$ we obtain the Einstein equation slightly modified by the NMC term:

\begin{equation}
(1+\kappa^{2}\xi\varphi^{2})G_{\mu\nu}=\kappa^{2}\tilde{T}_{\mu\nu}
\label{einstein}
\end{equation}

\noindent where we defined the stress-energy tensor $ T_{\mu\nu} $ such that:

\begin{equation}
\tilde{T}_{\mu\nu}=\nabla_{\mu}\varphi\nabla_{\nu}\varphi-\frac{1}{2}g_{\mu\nu}\nabla^{\gamma}\varphi\nabla_{\gamma}\varphi-Vg_{\mu\nu}-\xi
\bigg(g_{\mu\nu}\nabla(\varphi^{2})-\nabla_{\mu}\nabla_{\nu}(\varphi^{2})\bigg).
\label{tmunutilde}
\end{equation}

\noindent The familiar form of Einstein equations is clearly obtained in the limit of vanish $ \xi $, however there are two way to accommodate the coupling term to have those equations in their most familiar form.
 In a first approach, we can introduce an effective and $\varphi$-dependent gravitational constant as
\begin{equation}
G_{eff}\equiv\frac{G}{(1+\kappa^{2}\xi\varphi^{2})}
\end{equation}
so, Eq.~\eqref{einstein} can be re-written as
\begin{equation}
G_{\mu\nu}=\kappa_{eff}^{2}\tilde{T}_{\mu\nu}
\end{equation}
where $\kappa_{eff}^{2}\equiv8\pi G_{eff}$. In a second approach, we can consider a $\varphi$-independent gravitational
constant $G$ and accommodate the $\varphi$-dependence into the stress-energy tensor. Therefore, Eq.~\eqref{einstein} is
\begin{equation}
G_{\mu\nu}=\kappa^{2}{T}_{\mu\nu}
\end{equation}
\noindent where
\begin{equation}
T_{\mu\nu}\equiv\frac{\tilde{T}_{\mu\nu}}{(1+\kappa^{2}\xi\varphi^{2})}.
\label{tmunu}
\end{equation}

We can notice that both approaches produce two boundary values of scalar field for $\xi\rightarrow 0$:
\begin{equation}
\pm\varphi_{ crit}=\pm\frac{1}{\kappa\sqrt{|\xi|}}
\end{equation}
where the value of the inflaton field is divergent. Moreover, the conservation law for the energy-momentum tensor is valid for Eq.~\eqref{tmunu} due to the contracted Bianchi identities $\nabla^{\nu}G_{\mu\nu}=0$, while in Eq.~\eqref{tmunutilde}, is valid only in the case of $\varphi=const$.
Assuming a spatially flat FRW cosmology with line element $ds^{2}=g_{\mu\nu}dx^{\mu}dx^{\nu}=dt^{2}-a^{2}(t)\delta_{ij}dx^idx^j$, the Friedmann equations in the Jordan frame are:

\begin{align}
H^{2}=&\frac{\kappa^{2}}{3(1+\kappa^{2}\xi\varphi^{2})}\bigg[\frac{\dot{\varphi}^{2}}{2}+V(\varphi)-6\xi
H\varphi\dot{\varphi}\bigg]\label{friedmannI} \\
\frac{\ddot{a}}{a}=&-\frac{\kappa^{2}}{3(1+\kappa^{2}\xi\varphi^{2})}\bigg[\dot{\varphi}^{2}-V(\varphi)+3\xi
H\varphi\dot{\varphi}+3\xi\dot{\varphi}^{2}+3\xi\varphi\ddot{\varphi}\bigg]
\label{friedmannII}
\end{align}
According to the slow-roll approximation, the inflaton field slow rolls from the beginning to the end of inflation. The slow-roll conditions in the Hamilton-Jacobi formalism are expressed as [61]
\begin{equation}
\bigg|\frac{\ddot{\varphi}}{\dot{\varphi}}\bigg|\ll H,\quad\quad
\bigg|\frac{\dot{\varphi}}{\varphi}\bigg|\ll H,\quad\quad
\dot{\varphi}^{2}\ll V(\varphi),\quad\quad \bigg|\dot{H}\bigg|\ll
H^{2}.
\label{slowrollconditionham}
\end{equation}
The slow-roll parameters can be defined as
\begin{equation}
\epsilon\equiv\frac{-\dot{H}}{H^{2}},\quad\quad
\eta\equiv\frac{-\ddot{H}}{H\dot{H}},\quad\quad
\zeta\equiv\frac{V'\delta\phi}{\dot{\varphi}^{2}}=\frac{V'H}{2\pi\dot{\varphi}^{2}}
\end{equation}
where primes implies a derivative with respect to the inflaton field $\varphi$. We note that during the inflationary period, the slow-roll parameters remain less than unity and inflation does not end until the condition $\epsilon=1$ is met.

\subsection{Einstein frame}

The NMC term in the action written in the Jordan frame, Eq.~\eqref{actionj}, can be formally removed  considering the following conformal transformation to the Einstein frame:
\begin{equation}
\hat{g}_{\mu\nu}=\Omega^{2} g_{\mu\nu},
\end{equation}
with $\Omega^{2}=1+\kappa^{2}\xi\varphi^{2}$. The action in the Einstein frame takes therefore the form:

\begin{equation}
S_{E}=\int
d^{4}x\sqrt{-\hat{g}}\bigg(\frac{\hat{R}}{2\kappa^{2}}-\frac{1}{2}F^{2}(\varphi)\hat{g}^{\mu\nu}\partial_{\mu}\varphi\partial_{\nu}\varphi
-\hat{V}(\hat{\varphi})\bigg).
\label{actione}
\end{equation}

\noindent At the same time, the scalar field in this frame is defined as

\begin{equation}
F^{2}(\varphi)\equiv\bigg(\frac{d\hat{\varphi}}{d\varphi}\bigg)^{2}\equiv\frac{1+\kappa^{2}\xi\varphi^{2}({1}+6\xi)}{(1+\kappa^{2}\xi\varphi^{2})^{2}}
\end{equation}

\noindent and we deal with an effective potential

\begin{equation}
\hat{V}(\hat{\varphi})\equiv\frac{V(\varphi)}{(1+\kappa^{2}\xi\varphi^{2})^{2}}.
\end{equation}
When considering the Einstein frame, we need to transform our coordinate system using,
\begin{equation}
dT = \sqrt{\Omega}dt \qquad ; \qquad \hat{a} = \sqrt{\Omega}a
\end{equation}
to obtain the metric in the FRW form. By considering the FRW metric $d\hat{s}^{2}=dT^{2}-\hat{a}^{2}(T)\delta_{ij}dx^idx^j	$, the field equations in the Einstein frame take the form:
\begin{equation}
\hat{H}^{2}=\frac{\kappa^{2}}{3}\bigg[\frac{1}{2}\Big(\frac{d\hat{\varphi}}{dT}\Big)^{2}+\hat{V}(\hat{\varphi})\bigg],
\quad\quad\frac{\ddot{\hat{a}}}{\hat{a}}=-\frac{\kappa^{2}}{3}\bigg[\bigg(\frac{d\hat{\varphi}}{dT}\Big)^{2}-V(\varphi)\bigg],
\quad\quad\frac{d^{2}\hat{\varphi}}{dT^{2}}+3\hat{H}\frac{d\hat{\varphi}}{dT}+\frac{d\hat{V}}{d\hat{\varphi}}=0.
\end{equation}
Then the slow-roll conditions can straightforwardly written as,
\begin{equation}
\dot{\hat{\varphi}}^{2}\ll\hat{V},\quad\quad\ddot{\hat{\varphi}}\ll3\hat{H}\dot{\hat{\varphi}}
\label{slowrollcondition}
\end{equation}
and the slow-roll parameters are defined accordingly:
\begin{equation}
\hat{\epsilon}\equiv\frac{1}{2\kappa^{2}}\(\frac{\hat{V}^{\prime}(\hat{\varphi})}{\hat{V}(\hat{\varphi})}\)^{2},\quad
\hat{\eta}\equiv\frac{1}{\kappa^{2}}\(\frac{{\hat{V}}''(\hat{\varphi})}{\hat{V}(\hat{\varphi})}\),\quad
\hat{\zeta}\equiv\frac{1}{\kappa^{2}}
\(\frac{\hat{V}^{\prime}(\hat{\varphi}){\hat{V}}'''(\hat{\varphi})}{\hat{V}^{2}({\hat\varphi})}\)^{1/2}
\label{slowroll}
\end{equation}
where primes now imply a derivative with respect to the redefined scalar field $\hat{\varphi}$.


\end{document}